# Extraction of absorption coefficients from GaN nanowires grown on opaque substrates

*R. Jayaprakash[1,2], D. Ajagunna[2,3], S. Germanis[1], M. Androulidaki[2], K. Tsagaraki[2], A. Georgakilas[2,3], and \*N.T. Pelekanos[1,2]*

[1]Department of Materials Science and Technology, University of Crete, P.O. Box 2208, 71003 Heraklion, Greece

[2]Microelectronics Research Group, IESL-FORTH, P.O. Box 1385, 71110 Heraklion, Greece

[3]Department of Physics, University of Crete, P.O. Box 2208, 71003 Heraklion, Greece

Abstract

We demonstrate a new method to measure absorption coefficients in any family of nanowires, provided they are grown on a substrate having considerable difference in permittivity with the nanowire-air matrix. In the case of high crystal quality, strain-free GaN nanowires, grown on Si (111) substrates with a density of ~$10^{10}$ cm$^{-2}$, the extracted absorption coefficients do not exhibit any enhancement compared to bulk GaN values, unlike relevant claims in the literature. This may be attributed to the relatively small diameters, short heights, and high densities of our nanowire arrays.

N. T. Pelekanos

It is widely accepted that semiconductor nanowires (NWs) have a potential for low cost solar cell applications based on (1) their relaxed lattice matching requirements, due to strain accommodation at the NW free surface, providing flexibility in substrate selection and band-gap engineering, and (2) lesser material utilization, due to enhanced light absorption in NW arrays. On the latter very important issue, there has been extensive literature in recent years reporting on enhanced absorption characteristics of NW arrays, as a consequence of several physical effects such as their anti-reflective (AR) properties due to natural refractive index matching and grading[1–3], increased light trapping inside the NW array[3–5], and the excitation of resonant modes[6–10]. Most of the above literature refers to theoretical predictions based on solving wave propagation models in the NW array of increasing sophistication. Experimentally, increased absorption has been observed in direct transmission experiments on etched-down vertical Si nanowires placed on transparent[2] or semitransparent[3,5] substrates, and was attributed to reduced reflectance and enhanced light trapping effects in the NW array. The effect of light trapping was also used to interpret the enhanced performance of Si nanowire solar cells[4]. Regarding resonant effects, Cao et al[6] observed distinct resonances in the spectral response of single Ge nanowire photodetectors, that they attributed to the excitation of leaky mode resonances formed at the nanowire/air dielectric system. The excitation of resonant modes as a means to increase the absorption of NW arrays was discussed in several theoretical works[7–10] and was used to interpret the very high 13,8% photovoltaic efficiency of InP-based NW solar cells[11] as well as the single GaAs NW photovoltaic device with $\eta \approx 40\%$[12]. In spite of the importance of the issue, however, it is quite amazing that the works that have *directly* measured increased absorption of nanowire arrays have been very limited. This is mainly due to difficulties performing a transmission experiment through opaque substrates, as is the case for instance of GaAs NWs on Si or GaAs substrates, of GaN NWs grown on Si substrates or GaN templates, or of InP NWs homoepitaxially grown on InP substrates. In this work, we demonstrate that it is possible to extract absorption coefficients from GaN NWs grown on Si<111> substrates, by analyzing merely reflectivity spectra. This method can be extended to any family of NWs, provided they are grown on a substrate having considerable difference in permittivity with the nanowire-air matrix.

The GaN NWs studied here are grown along the [0001] direction on a Si (111)-oriented substrate, by plasma-assisted molecular beam epitaxy (PAMBE), at nitrogen rich conditions. Several samples were grown at variable conditions, having similar diameters, heights and densities. The Scanning Electron Microscopy (SEM) top and lateral view for one of the samples (G1643) is shown in Fig. 1. The GaN NWs are upright and quite uniform, having average heights (*h*) of 450nm and diameters (*d*) of 24nm. A good fraction of the NWs possess pitted tips, others stepped surfaces, while the rest flat top surfaces. The pitted tip of nanowires is attributed to inversion domain boundaries while the stepped surface is due to prismatic stacking faults or planar defects[13]. A separate two-dimensional (2D) GaN thin film used for modeling is grown using the same technique along the [0001] direction on a Si substrate (111). The thickness of the 2D film is approximately 810nm and possesses good crystalline quality. Most of the experimental results presented in this paper are based on sample G1643, even though very similar results were obtained on all of the samples available to us.

Photoluminescence (PL) and reflectivity (RFL) measurements are made on an Acton, 0.5-meter focal length, triple-grating, imaging Spectrograph, with a resolution of 0.02nm at 435.8nm and an accuracy of ±0.02nm. A Kimmon IK series, continuous wave, He-Cd laser is used to excite the sample at 325nm, for PL measurements. A Xenon lamp, in a high precision

Hamamatsu lamp housing, is used as a source for RFL measurements. The incident beam is focused on to the sample placed in an APD helium closed-circuit cryostat, typically at an angle of incidence of ≈15°, and the PL/RFL signal is focused onto the spectrometer (setup 1). In this setup, only specular reflectance can be measured. To correct for the lamp's spectral response, a Newport 5108 UV-Vis mirror is used as a reference, having a nearly flat response and a reflectivity of ~90%, in the range of 300-600nm. The sample temperature can be varied from 20K to 310K using a Scientific Instruments, series 5500 temperature controller. All high-resolution PL/RFL spectra are taken using a 2400g/mm holographic UV grating, while the lower resolution spectra are taken using a 600g/mm grating, blazed at 300nm. The absorption coefficients reported in the paper depend strongly on a precise estimate of the filling factor of the NW array. This was determined accurately by simulating the Fabry Perot region, of the *total* reflectivity spectrum, (i.e. *specular + diffuse*), acquired with a 150mm integrating sphere, equipped on a Perkin Elmer LAMBDA 950 UV/Vis/NIR Spectrophotometer. In this setup 2, apart from the total reflectivity, the diffuse reflectivity is separately recorded.

The GaN NWs are of high optical quality, as testified by comparing PL and RFL spectra in Fig. 2a,b as a function of temperature. In the low temperature PL, the A and B free excitons are clearly visible at 3.4782±0.0002 eV and 3.4825±0.0005 eV, with the Donor Bound Exciton (DBE) emission placed approximately 7 meV below the A exciton, at 3.4713±0.0002 eV. These positions are in excellent agreement with those reported for GaN films grown on bulk GaN substrates[14,15], highlighting the absence of any residual strain in the nanowires. The PL spectrum is dominated by a double structure around 3.45 eV, which is very usual in N-polar GaN films, GaN samples with columnar structure, as well as in self-induced GaN nanowires. Its origin has been a topic of debate[16–18]: several groups have reported this doublet, as due to excitons bound to inversion domain boundaries[16,19], while Corfdir et al[20] have attributed this peak to two electron satellite (TES) of donor bound excitons, close to the surface. A relatively weak peak around 3.42 eV can be attributed to excitons bound to surface defects[16] or to stacking faults[16,17]. With increasing temperature, the DBE emission dies quickly, becoming weaker than the free exciton lines, already at ≈60K (see Suppl. Information). The A and B excitons start merging around 100K, and a weak shoulder appears on the higher energy side, 23meV from the A exciton, an energy position reported for C exciton in bulk GaN film[14,15]. The defect-related doublet structure around 3.45eV persists up to 200K, above which the PL spectrum is dominated by free exciton emission, a clear sign of high optical quality. Finally at room temperature, the A, B and C excitonic peaks merge into a single PL peak centered at 3.422±0.001 eV, similar to bulk GaN films.

In order to confirm the excitonic positions and probe their oscillator strengths, specular RFL measurements are undergone, using an angle of incidence of ≈15°, as shown in Figs. 2a,b. At low temperature, the A and B excitons are clearly visible, perfectly corresponding to their PL positions. We also observe a third peak at 3.5007±0.0005 eV, matching well with the position of C excitons reported in bulk GaN films[14,15]. The A and B excitons appear to have very similar oscillator strengths, which is consistent with the picture of strain-free nanowires, considering that in strained GaN films, the B excitons tend to be stronger than the A ones[21,22]. Interestingly, the oscillator strength of the C exciton is comparable to that of A and B excitons. This is somewhat puzzling, considering that in small angle RFL experiments on c-axis oriented GaN films, the C-excitons are strongly suppressed[14,15] in accordance with the well-known polarization selection rules for GaN[23], favoring the A and B excitons for polarization perpendicular to the c-axis, and the C excitons for polarization parallel to it. Possible explanation for this is that, although the majority of NWs seem upright in Fig.1, there is a scarce population of tilted NWs, or small deviations from verticality, relaxing the selection rule and enhancing the C exciton signature, in conjunction with the well-known anisotropy of the nanowires amplifying the absorption cross section for polarizations parallel to the c-axis[24]. In order to ensure that the 3.5eV peak behaves indeed as a C exciton, a polarization-dependent RFL experiment is carried out at large incidence angles (~65°), using a Glan Thomson polarizer. Unlike the small angles, where all polarization components are essentially perpendicular to the c-axis, the large angles serve the purpose of creating significant parallel and perpendicular to the c-axis polarization components depending on the polarizer angle. The results are as shown in Fig. 2c. The angle 0º corresponds to p-polarization whereas the angle 90º to s-polarization. As expected, the C exciton line is much stronger at angle 0º, where a significant polarization component parallel to the c-axis exists, whereas is significantly reduced at angle 90º, where all polarization components are perpendicular to c-axis. With increasing temperature, the A, B and C exciton features in the RFL spectra broaden and progressively merge into a single excitonic peak at 295K, perfectly matching with the PL emission peak as depicted in Fig. 2b.

However, what is most interesting in the optical characterization of our NW samples is that the RFL spectrum looks very much like transmission. This is quite evident in the wide range RFL spectrum of Fig. 2d, where we clearly distinguish the characteristics of a transmission spectrum, with the low absorption region below gap and the presence of exciton absorption peaks at the A, B and C exciton positions[23]. Please note that the A and B excitons appear here merged into a single peak due to the lower resolution grating used. There is a very simple picture allowing us to understand the "transmission-like" characteristics of the RFL spectrum. If we assume negligible reflectivity at the top GaN NW/air interface, then the specularly reflected

N. T. Pelekanos

spectrum of Fig.2d derives from light reflected at the bottom GaN NW/Si interface, after having traversed twice the GaN NW array, containing thus important information about the absorption characteristics of the NW array. In the rest of the paper, we validate this simple picture and use it to extract for the first time useful absorption coefficients for a NW system grown on an opaque substrate.

A key observation, strongly suggesting the anti-reflective character of the GaN NW/air interface, is the lack of any pronounced Fabry-Perot oscillation in the transparent region of GaN, as shown in Fig. 3. By contrast, very intense Fabry-Perot oscillations are observed in the total reflectivity spectrum of the reference GaN/Si thin film, shown in Fig. 4a, consistent with having two optically-flat top and bottom interfaces. As we show next, by simulating the Fabry-Perot region of the total reflectivity spectrum for the two types of samples, we are able to extract quantitative information, and demonstrate the very low reflectivity values of the top GaN NW surface. Towards this end, it is necessary to perform some high precision total reflectivity measurements, taking into account the possible contribution of diffuse reflectivity. Thus, all measurements essential for modeling are taken with the help of setup 2 in two configurations: 1) with specular white plate, 2) with specular light trap, as shown in the schematic of Fig. 3. The angle of incidence is 8°. With the first configuration, we measure the total reflectivity of the nanowires, whereas with the second, the specular reflection is trapped and we are led to accurate measurement of the diffuse reflectivity. Subtraction of the diffuse from the total reflectivity gives the specular reflectivity, as shown in Fig. 3. The shape of the specular reflectivity is very similar to the shape returned by measurements made with setup 1. In addition, the diffuse reflectivity of the nanowires appears to be significant even in the long wavelength region where it amounts to ≈5%. With decreasing wavelength, the NW diameter to wavelength ratio increases[25] and the

N. T. Pelekanos

diffuse reflectivity becomes more prominent, reaching values over 20% near the GaN gap. Above the gap, the diffuse signal drops down sharply, due to the abrupt onset of absorption, limiting the volume for diffuse scattering to the very top part of the NW array. We interpret the 8.5% plateau at short wavelengths as due to surface diffuse scattering. Overall, however, it is clear from Fig.3 that the diffuse reflectivity cannot be neglected, making it essential to compare our modeling results with total reflectivity spectra, rather than specular alone.

In order to simulate the reflectivity in the transparent region for the two types of structures a transfer matrix method is utilized[26], as described in detail in the supplementary information. Several assumptions go along with this model: (1) the film and substrate are optically isotropic and homogeneous, (2) the surface of the film is flat and smooth, (3) there is no residual absorption in the Fabry-Perot region, and (4) the back reflection from the substrate/air bottom interface can be neglected, which applies in our case since the Si substrate is strongly absorbing in the wavelength range under consideration. According to this model, the power reflectivity is calculated from an effective characteristic matrix $M$ of the multilayer system, constructed by multiplication of the characteristic matrices of the individual layers. The model is tested first, simulating the reflectivity spectrum of the GaN/Si thin film with nominal thickness of about 810 nm. All the initial assumptions made for the model are applicable here. The refractive index $n$ for GaN and Si is calculated according to the Sellmeier equation, with the coefficients taken from ref.[27]. The GaN layer thickness is used as the only adjustable parameter. The incidence angle is taken 8° as in the integrating sphere setup. As can be seen in Fig.4a, the simulation produces excellent agreement with the experimental curve, for a thickness of 808.4nm, which is very close to the nominal thickness of the film.

Next, we apply the transfer matrix model to our GaN nanowires on Si, which have an average height of approximately 450nm (from SEM lateral view). Obviously, the NW array does not meet the first two assumptions of the model, i.e. they are not optically isotropic and homogeneous, nor do they have a flat and smooth surface. Instead, the nanowires can be compared to a set of anisotropic scatterers dispersed in a medium of air forming a dielectric matrix that is not homogeneous[24]. In order to turn the NW array into a homogeneous medium, with a flat and smooth surface, an effective medium theory (EMT)[24,28,29] has to be applied. EMT calculates an effective permittivity for the dielectric matrix provided the individual permittivities $\varepsilon(\lambda) = n^2$, hereafter denoted as $\varepsilon$, and the volume fractions of the constituents are known[29]. However, a prerequisite for the validity of EMT is that the wavelength of incident light is much larger than the nanowire average diameters and interspacings (long-wavelength limit). This holds in our case, since the wavelength range of interest is between 400 nm and 845 nm, which is much larger than the nanowire average diameter ($d$=24nm) as well as the spacing between them, which based on the volume fractions reported on Table I, vary in the range 50-150 nm.

Considering that the nanowires are inherently anisotropic scatterers[24], in the sense that light polarized parallel to the NW long axis is scattered more efficiently than light polarized perpendicular to it, any effective medium theory homogenizing the NW array, should take into account this anisotropy by using an effective permittivity tensor (see Suppl. Information). In our configuration, however, this is not necessary, because the NWs are oriented along the c-axis, and the angle of incidence is near to normal, implying that both $s$ and $p$ polarizations of the incident light are practically perpendicular to the nanowire axis. As such, the medium's response is expected to be isotropic and can be described by a single effective ordinary permittivity.

There are two well-known effective medium theories for calculating the effective ordinary permittivity of a dielectric matrix, based on Lorentz-Lorentz Clausius-Mossotti equation[29]: (1) the Maxwell Garnett theory (MGT), and the (2) Bruggeman effective medium approximation (EMA)[28,29]. Between the two, the MGT is less adapted to describe a GaN NW system, due to the large refractive index contrast between air and GaN, condition for which MGT fails. Therefore, hereafter, we apply the Bruggeman EMA, for which it is essential to know the permittivities of air and GaN, as well as the volume fraction of the nanowires in the dielectric matrix. To get an idea about the NW volume fraction, we take a closer look at the SEM images, where one can easily realize that although most of the nanowires are upright, they have different heights, and as mentioned earlier, a fraction of them have stepped surfaces while some are pitted towards the top. In other words, the SEM images suggest that the effective ordinary permittivity should be graded along the height of the dielectric matrix. Accordingly, we divide for simplicity the dielectric matrix into two layers, as depicted in Fig. 4d, introducing four parameters: $h_1$-height of top layer, $f_1$-volume fraction of top layer, $h_2$-height of bottom layer and $f_2$-volume fraction of bottom layer. The minimum height of most of the nanowires is determined by SEM lateral view to be 345.5nm and is assigned to $h_2$. In order to estimate volume fractions, we rely on a grey scale top view SEM image of the nanowires as shown in Fig. 1. Each pixel on the image has an intensity corresponding to the height of the nanowire in that particular region. Next, we convert the grey scale image into black and white (B/W) image using 2 thresholds: (1) by keeping the threshold low, such that the B/W image contains information only about the nanowires really close to the surface, and (2) by keeping the threshold high such that the B/W image contains information even about nanowires having the minimum height of 345.5nm. The bright regions of the B/W image correspond to nanowires and the dark regions to air. The ratio of

N. T. Pelekanos

bright pixels to total pixels gives us the volume fraction of the nanowires. The low threshold B/W image gives us an estimate of $f_1$ to be 0.108 in the particular image of Fig.1, and the high threshold image gives us an estimate of $f_2$ to be 0.1919. These values of $f_1$, $h_2$ and $f_2$ are used in the simulation, keeping $h_1$ as a varying parameter to account for the irregular heights, stepped surface and pitted nature of the NW array towards the top.

In the Bruggeman EMA[29], the effective permittivity $\varepsilon$ of the dielectric matrix is given by:

$$f\left(\frac{\varepsilon_n-\varepsilon}{\varepsilon_n+K\varepsilon}\right) + (1-f)\left(\frac{\varepsilon_m-\varepsilon}{\varepsilon_m+K\varepsilon}\right) = 0 \quad (1)$$

, where $\varepsilon_n$ is the permittivity of the nanostructure, $\varepsilon_m$ the permittivity of the surrounding matrix, $f$ the volume fraction of the nanostructures, and $K=1$ for an array of cylinders arranged such that the incident beam is collinear to their long axis. Substituting for $f_1$ and $f_2$ in equation (1), and using the GaN and air permittivities, we estimate the effective ordinary permittivity $\varepsilon$ in each of the two layers, based on which we can calculate the characteristic transfer matrix of the two-layer system and solve for the total reflectivity according to the transfer matrix model, keeping $h_1$ as the sole adjustable parameter. The simulation precisely reproduces the experimental data of Fig.4b for $h_1$=120.1nm. The total average height of the nanowires based on the simulation turns out to be $(h_1+h_2)$ = 465.6nm, very close to the average height of the nanowires ~450nm estimated from SEM lateral view. Using $n = \sqrt{\varepsilon}$ as well as the basic relation for the power reflectivity at a given interface,

$$R = \left(\frac{n-n'}{n+n'}\right)^2 \quad (2)$$

with $n$ and $n'$ being the refractive indices above and below the interface, an additional plot is made in Fig. 4c for the top, middle and bottom power reflectivities of the GaN NW/air matrix, demonstrating that the top and middle reflectivities are negligible (< 0.2%), while on the other hand the bottom reflectivity at the interface with Si is as high as 35%. These numbers strongly support our assertion that light incident on the NW/air matrix hardly gets reflected from the nanowires, but instead most of it gets transmitted through them, suffers reflection at the bottom interface with Si, and gets transmitted back again, thereby travelling a distance equivalent to twice the average height ($h$) of the nanowires, i.e. 931.2nm for the sample of Fig.1. Thus, a modified Beer-Lambert's law can be applied for the specular reflectivity:

$$I = I_0 \cdot (1 - D_S)^2 \cdot R_0 \cdot e^{-(\alpha + D_V)d} \qquad (3)$$

, where $I_0$ is the intensity of incident light, $D_S$ and $D_V$ represent diffuse scattering losses at the surface and the volume of the NW array, respectively, $R_0$ is the reflectivity at the NW array/Si interface, $\alpha$ the absorption coefficient of the NW array, and $d$ is the path length (=$2h$). Assuming for a moment that $D_S = D_V = 0$, and considering a spectrally flat $R_0$, we can estimate ($I_0 \cdot R_0$) from the reflectivity signal just below the band gap of GaN, and turn a specular reflectivity spectrum, such as in Fig.2, into a quantitative absorption spectrum. In Fig. 5a, we plot a set of temperature-dependent absorption coefficient spectra, obtained as described above, from specular reflectivity spectra between 25K and 295K. The spectra are upshifted between them by 4,000cm$^{-1}$, for clarity. At low temperatures, we observe the prominent A, B and C excitonic features as two separate peaks, one for the A and B excitons merged together, and one for the C exciton. With increasing temperature, the two peaks progressively merge into a single excitonic peak containing contributions from all A, B and C excitons. About 100meV above the A exciton, we distinguish a characteristic bump-like feature which is attributed to exciton-LO phonon bound states, as previously reported for bulk GaN[30,31]. This bump is more visible in Fig. 5b, where we take a closer look at the 25K and 295K absorption spectra, and its assignment is further supported by the fact that with increasing temperature it follows the trails of the other excitonic peaks, confirming thus its excitonic character.

We discuss now the amplitude of the absorption coefficients $\alpha$ extracted as a function of temperature between 25K and 295K. At low temperature, the $\alpha$-values refer to the maximum of the merged A and B exciton peak, while at high temperature to the single merged A, B and C excitonic peak. From Fig.5a, $\alpha$ is equal to 13,000 cm$^{-1}$ at 25K and to about 10,500 cm$^{-1}$ at 295K. These values can be corrected to about 17,000 cm$^{-1}$ at 25K and 14,500 cm$^{-1}$ at 295K, by considering the spectral dependence of diffuse scattering discussed in Fig. 3. The correction is achieved by making use of equation 3, and assuming that $D_V$ is zero above the gap, while $D_S$ remains relatively constant in the 300-400nm range of interest and is equal to 8.5%, as seen from Fig.3. To compare now with the respective $\alpha$-values reported in the literature, which are about 140,000 cm$^{-1}$ at 77K and 95,000 cm$^{-1}$ at 295K for bulk GaN films[30], we further need to take into account the deficit amount of material available for absorption in the NW array, compared to a thin film. Accordingly, we estimate the film factor $F$, which is the volume fraction of available GaN material in the NW array, given by:

$$F = \frac{h_1+h_2}{h_1 f_1 + h_2 f_2} \qquad (4)$$

Substituting in equation (4) the values of $h_1$ returned by the simulation, and the values of $h_2, f_1$ and $f_2$ determined by SEM, we estimate a film factor of 5.88. Thus, the corrected absorption coefficient is given by:

$$\alpha_{corrected} = \alpha_{experimental} \times F \qquad (5)$$

Using equation (5), the corrected $\alpha$-values are estimated to be ~100,000cm$^{-1}$ at 25K and ~85,000cm$^{-1}$ at 295K, as shown in Fig. 5b.

N. T. Pelekanos

The corrected absorption coefficients are still smaller than the values reported by Muth et al.[30] for GaN thin films, in spite of prior claims in the literature for enhanced absorption of NW arrays, due to AR property, light trapping effects, and excitation of resonant guided modes. The AR property of the GaN NWs is confirmed in our work by the very low reflectivity values (<0.2%), estimated for the top NW/air interface following the discussion of Fig. 4. However, this AR property should not contribute to the deduced absorption coefficients, as it has been explicitly taken into account in the analysis leading to them. In addition, light trapping effects are not favored in the rather short ($h \approx 450$nm) and dense ($3 \cdot 10^{10}$ cm$^{-2}$) NW arrays of this work. Similarly, resonant mode effects should be negligible in our case of very thin ($d \approx 24$nm) nanowires. In other words, not much of an enhancement is expected for the NW array at hand, anyway. On the other hand, the smaller, compared to bulk, $\alpha$-values could be possibly attributed to the large surface to volume ratio of nanowires and the presence of surface defects. In GaN NWs, surface states are formed by Ga or N dangling bonds, binding impurities on the surface[32]. These surface states have been reported to cause Fermi level pinning[32–35], drastically affecting the optical and electrical properties. Pfuller et.al[18] showed an enhancement in PL by continuous UV exposure for a certain time, that they attributed to a photo-induced desorption of oxygen from the NW sidewalls, unpinning the Fermi level and enhancing the PL efficiency. Tuoc et.al[34] showed that core shell nanowires perform better than bulk nanowires alone, again confirming the adverse influence of surface defects in GaN nanowires. Evidence for surface defects in our nanowires is brought about by the PL peak at 3.42eV, which is typically attributed to excitons bound at surface defects[16]. Based on the above, it is quite reasonable to suggest that 1-2 nm from the GaN surface, do not contribute to the absorption process. Assuming just 1 nm of "dead" shell is sufficient to modify the film factor from 5.88 to 7, and increase the $\alpha$-values to 120,000cm$^{-1}$ at

N. T. Pelekanos

25K and to 100,000cm$^{-1}$ at 295K, i.e. quite similar to the values reported by Muth et al. for bulk GaN.

In summary, we introduce here a new method to extract absorption coefficients from nanowire arrays grown on opaque substrates. The method is based on simple reflectivity measurements on the as-grown samples, and can be extended to any family of nanowires, provided they are grown on substrates with sufficient permittivity difference compared to the nanowire-air matrix. In the studied case of high optical quality, strain-free GaN nanowires, we observe strong excitonic features, both in the emission and absorption coefficient spectra, all the way up to room temperature. However, we find no evidence of enhancement in the amplitude of the absorption coefficients, at least compared to bulk GaN. This could be attributed to the particular structural characteristics of the studied nanowire arrays, not favouring light trapping effects, nor the excitation of resonant modes. Ongoing experiments focus on variable nanowire samples, prone to show absorption enhancement effects.

N. T. Pelekanos

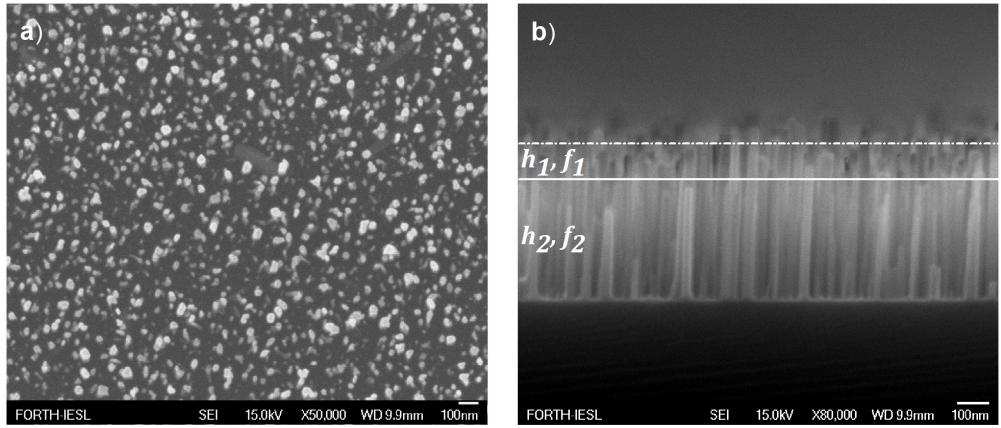

**Figure 1.** (a) SEM top view and (b) SEM lateral view of GaN (0001) nanowires on Si (111).

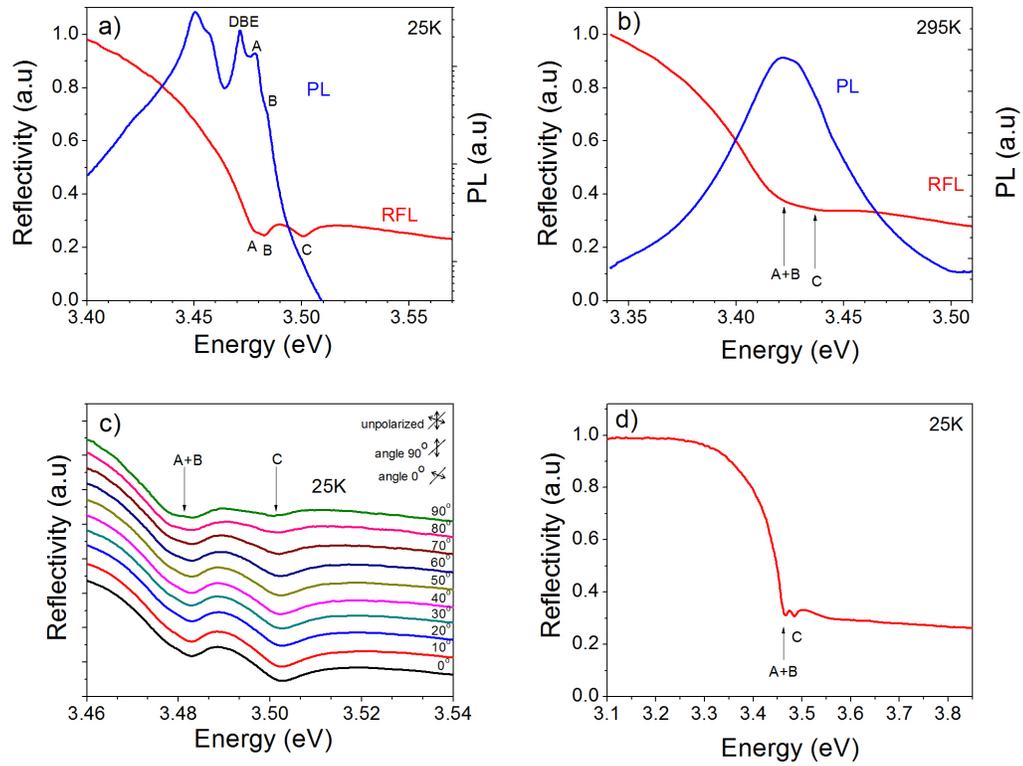

**Figure 2.** (a) Comparison of PL (blue) and RFL (red) from GaN NWs at LT, (b) Comparison of PL (blue) and RFL (red) from GaN NWs at RT, (c) Polarization-resolved RFL at a large angle of incidence of 65°, demonstrating polarization dependence of the excitons, by varying the polarizer angle from 0 to 90°, (d) Wide range RFL spectrum of GaN NWs showing transmission-like characteristics.

N. T. Pelekanos

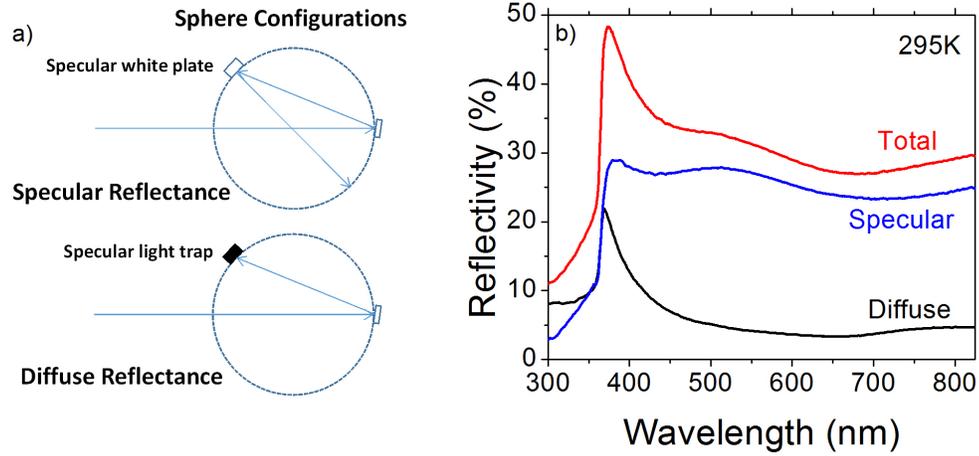

**Figure 3.** (a) Different integrating sphere configurations to measure the total and diffuse reflectivity in setup 2. (b) Total (red), specular (blue) and diffuse (black) reflectivity spectra obtained from the GaN NW sample, at room temperature.

N. T. Pelekanos

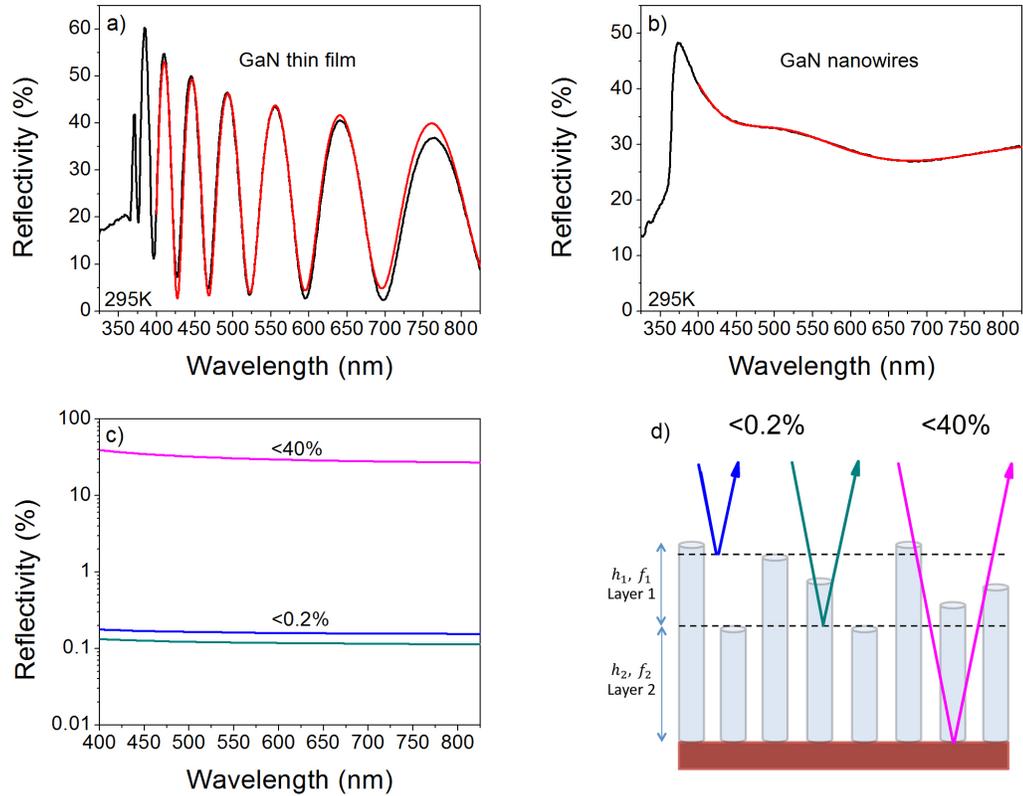

**Figure 4.** Simulation of the total RFL spectrum of (a) a GaN/Si thin film, and (b) of GaN NWs grown on Si. (c) Deduced power reflectivities from the simulation, at different interfaces of the GaN NW array, such as top (blue), middle (green) and bottom interface (purple). (d) An illustrative diagram showing the power reflectivities from the different GaN NW interfaces.

N. T. Pelekanos

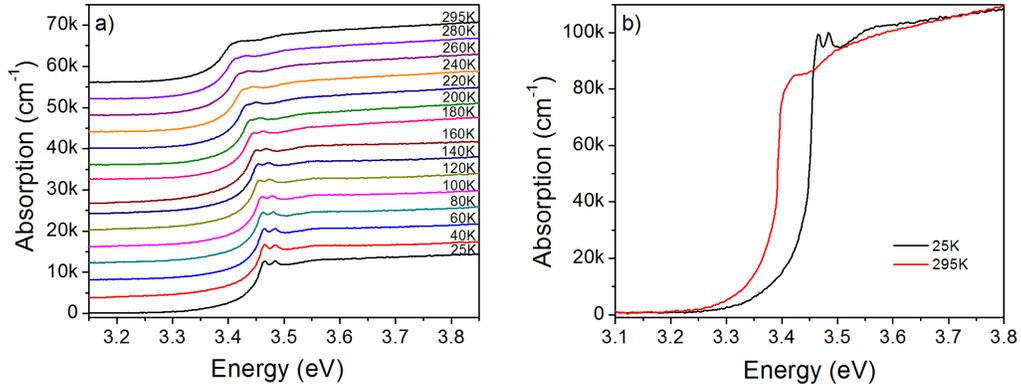

**Figure 5.** (a) Absorption spectra from GaN NWs with varying temperature between 25K and 295K. (b) Low and room temperature absorption spectra, emphasizing the exciton absorption lines, along with the phonon assisted absorption peak.

AUTHOR INFORMATION

**Corresponding Author**

N. T. Pelekanos

Department of Materials Science and Technology, University of Crete, P.O. Box 2208, 71003 Heraklion, Greece / Microelectronics Research Group, IESL-FORTH, P.O. Box 1385, 71110 Heraklion, Greece.

Email: pelekano@materials.uoc.gr

Phone: +30-2810 394107

**Author Contributions**

The manuscript was written through contributions of all authors. All authors have given approval to the final version of the manuscript.

**Funding Sources**

This work was supported by the European Initial Training Network "ICARUS" and the European Social Fund and National resources through the THALES program "NANOPHOS", and the ARISTEIA program "NILES".

N. T. Pelekanos

ABBREVIATIONS

Ga Gallium, N Nitrogen, Si Silicon, Ge Germanium, GaN Gallium Nitride, InP Indium Phosphide, GaAs Gallium Arsenide, N-polar Nitrogen-polar, eV electron Volt, meV milli electron Volt, NW Nanowire, NWs Nanowires, PL Photoluminescence, RFL Reflectivity, AR Anti-Reflective, TES Two Electron Satellite, EMT Effective Medium Theory, MGT Maxwell Garnett Theory, EMA Effective Medium Approximation, PAMBE Plasma Assisted Molecular Beam Epitaxy, SEM Scanning Electron Microscopy, TEM Transmission Electron Microscopy, B/W Black and White, UV Ultraviolet, Vis Visible, NIR Near Infrared

N. T. Pelekanos

References

(1) Hu, L.; Chen, G. *Nano Lett*. **2007**, *7*, 3249–52.

(2) Zhu, J.; Yu, Z.; Burkhard, G. F.; Hsu, C.-M.; Connor, S. T.; Xu, Y.; Wang, Q.; McGehee, M.; Fan, S.; Cui, Y. *Nano Lett*. **2009**, *9*, 279–82.

(3) Tsakalakos, L.; Balch, J.; Fronheiser, J.; Shih, M.-Y.; LeBoeuf, S. F.; Pietrzykowski, M.; Codella, P. J.; Korevaar, B. a.; Sulima, O. V.; Rand, J.; Davuluru, A.; Rapol, U. *J. Nanophotonics* **2007**, *1*, 013552.

(4) Garnett, E.; Yang, P. *Nano Lett*. **2010**, *10*, 1082–7.

(5) Lu, Y.; Lal, A. *Nano Lett*. **2010**, *10*, 4651–4656.

(6) Cao, L.; White, J. S.; Park, J.-S.; Schuller, J. a; Clemens, B. M.; Brongersma, M. L. *Nat. Mater*. **2009**, *8*, 643–7.

(7) Lin, C.; Povinelli, M. L. *Opt. Express* **2009**, *17*, 19371–81.

(8) Kupec, J.; Stoop, R. L.; Witzigmann, B. *Opt. Express* **2010**, *18*, 27589–27605.

(9) Kailuweit, P.; Peters, M.; Leene, J.; Mergenthaler, K.; Dimroth, F.; Bett, A. W. *Prog. Photovoltaics Res. Appl*. **2012**, *20*, 945–953.

(10) Guo, H.; Wen, L.; Li, X.; Zhao, Z.; Wang, Y. *Nanoscale Res. Lett*. **2011**, *6*, 617.

(11) Wallentin, J.; Anttu, N.; Asoli, D.; Huffman, M.; Aberg, I.; Magnusson, M. H.; Siefer, G.; Fuss-Kailuweit, P.; Dimroth, F.; Witzigmann, B.; Xu, H. Q.; Samuelson, L.; Deppert, K.; Borgström, M. T. *Science* **2013**, *339*, 1057–60.

N. T. Pelekanos

(12) Krogstrup, P.; Jørgensen, H. I.; Heiss, M.; Demichel, O.; Holm, J. V; Aagesen, M.; Nygard, J.; Fontcuberta i Morral, A. *Nat. Photonics* **2013**, *7*, 306.

(13) Cherns, D.; Meshi, L.; Griffiths, I.; Khongphetsak, S.; Novikov, S. V.; Farley, N. R. S.; Campion, R. P.; Foxon, C. T. *Appl. Phys. Lett.* **2008**, *93*, 111911.

(14) Kornitzer, K.; Ebner, T.; Grehl, M.; Thonke, K.; Sauer, R.; Kirchner, C.; Schwegler, V.; Kamp, M.; Leszczynski, M.; Grzegory, I.; Porowski, S. *Phys. status solidi* **1999**, *216*, 5–9.

(15) Monemar, B. *J. Phys. Condens. Matter* **2001**, *13*, 7011–7026.

(16) Reshchikov, M. a.; Huang, D.; Yun, F.; Visconti, P.; He, L.; Morkoç, H.; Jasinski, J.; Liliental-Weber, Z.; Molnar, R. J.; Park, S. S.; Lee, K. Y. *J. Appl. Phys.* **2003**, *94*, 5623–5632.

(17) Geelhaar, L.; Ch, C.; Jenichen, B.; Brandt, O.; Pf, C.; Steffen, M.; Rothemund, R.; Reitzenstein, S.; Forchel, A.; Kehagias, T.; Komninou, P.; Dimitrakopulos, G. P.; Karakostas, T.; Lari, L.; Chalker, P. R.; Gass, M. H.; Riechert, H. *Quantum* **2011**, *17*, 878–888.

(18) Pfüller, C.; Brandt, O.; Grosse, F.; Flissikowski, T.; Chèze, C.; Consonni, V.; Geelhaar, L.; Grahn, H.; Riechert, H. *Phys. Rev. B* **2010**, *82*, 1–5.

(19) Schuck, P. J.; Mason, M. D.; Grober, R. D.; Ambacher, O.; Lima, a. P.; Miskys, C.; Dimitrov, R.; Stutzmann, M. *Appl. Phys. Lett.* **2001**, *79*, 952.

(20) Corfdir, P.; Lefebvre, P.; Ristić, J.; Valvin, P.; Calleja, E.; Trampert, a.; Ganière, J.-D.; Deveaud-Plédran, B. *J. Appl. Phys.* **2009**, *105*, 013113.

N. T. Pelekanos

(21) Shan, W.; Fischer, a. J.; Hwang, S. J.; Little, B. D.; Hauenstein, R. J.; Xie, X. C.; Song, J. J.; Kim, D. S.; Goldenberg, B.; Horning, R.; Krishnankutty, S.; Perry, W. G.; Bremser, M. D.; Davis, R. F. *J. Appl. Phys.* **1998**, *83*, 455–461.

(22) Fischer, a. J.; Shan, W.; Song, J. J.; Chang, Y. C.; Horning, R.; Goldenberg, B. *Appl. Phys. Lett.* **1997**, *71*, 1981–1983.

(23) Gil, B. In *Group III nitride semiconductor compounds: physics and applications*; Clarendon Press, 1998; pp. 158–181.

(24) Gomez Rivas, J.; Muskens, O.; Borgstrom, M. T.; Diedenhofen, S.; Bakkers, E. In *One-Dimensional Nanostructures*; Springer New York, 2008; pp. 127–145.

(25) Muskens, O. L.; Rivas, J. G.; Algra, R. E.; Bakkers, E. P. A. M.; Lagendijk, A. *Nano Lett.* **2008**, *8*, 2638–2642.

(26) Furman, S. A.; Tikhonravov, A. V. In *Basics of optics of multilayer systems*; Atlantica Séguier Frontières, 1992; Vol. 0, pp. 1–102.

(27) Bass, M.; DeCusatis, C.; Enoch, J.; Li, G.; Mahajan, V. N.; Lakshminarayanan, V.; Stryland, E. Van; MacDonald, C. *Handbook of Optics: Optical properties of materials, nonlinear optics, quantum optics, Volume 4*; McGraw-Hill Prof Med/Tech, 2009; p. 1152.

(28) Sihvola, A. H.; Pekonen, O. P. M. *J. Phys. D. Appl. Phys.* **1996**, *29*, 514–521.

(29) Baxter, J. B.; Schmuttenmaer, C. a *J. Phys. Chem. B* **2006**, *110*, 25229–39.

(30) Muth, J. F.; Lee, J. H.; Shmagin, I. K.; Kolbas, R. M.; Casey, H. C.; Keller, B. P.; Mishra, U. K.; DenBaars, S. P. *Appl. Phys. Lett.* **1997**, *71*, 2572.

N. T. Pelekanos

(31) Gil, B. In *Group III nitride semiconductor compounds: physics and applications*; 1998; pp. 182–241.

(32) Armstrong, A.; Li, Q.; Lin, Y.; Talin, a. a.; Wang, G. T. *Appl. Phys. Lett.* **2010**, *96*, 163106.

(33) Calarco, R.; Marso, M.; Richter, T.; Aykanat, A. I.; Meijers, R.; V D Hart, A.; Stoica, T.; Lüth, H. *Nano Lett.* **2005**, *5*, 981–4.

(34) Tuoc, V. N.; Huan, T. D.; Lien, L. T. H. *Phys. Status Solidi* **2012**, *9*, 1–9.

(35) Bertness, K. A.; Member, S.; Sanford, N. A.; Davydov, A. V *Quantum* **2011**, *17*, 847–858.

N. T. Pelekanos